\pdfoutput=1
\documentclass[a4paper,11pt]{article}
\usepackage{amsopn}
\usepackage{amsmath}
\usepackage{amssymb}
\usepackage{hyperref}
\usepackage{subfig}
\usepackage{bbm}
\usepackage{authblk}
\usepackage{cite}
\usepackage{graphics}
\usepackage{subfig}
\usepackage[braket, qm]{qcircuit}

\newcommand{\ketbra}[2]{\ensuremath{\ket{#1}\bra{#2}}}

\newcommand{\XX}{\mathcal{X}}
\newcommand{\YY}{\mathcal{Y}}
\newcommand{\ZZ}{\mathcal{Z}}

\newcommand{\LL}{\mathcal{L}}
\newcommand{\UU}{\mathcal{U}}
\newcommand{\DD}{\mathcal{D}}

\newcommand{\Tr}{\mathrm{Tr}}
\newcommand{\tr}{\Tr}
\newcommand{\Id}{\mathbbm{1}}
\newcommand{\ii}{\mathrm{i}}

\newcommand{\ee}{\mathrm{e}}
\newcommand{\proj}[1]{\ketbra{#1}{#1}}

\newtheorem{theorem}{Theorem}
\newtheorem{definition}{Definition}
\newtheorem{remark}{Remark}

\newtheorem{corollary}{Corollary}

\newenvironment{theproof}[1][Proof]{\noindent\textbf{#1.} }{\ 
\hfill\rule{0.5em}{0.5em}}

\title{Quantifying channels output similarity with applications to quantum 
control}

\author[1]{{\L}ukasz Pawela\thanks{lpawela@iitis.pl}}
\author[1,2]{Zbigniew Pucha{\l}a\thanks{z.puchala@iitis.pl}}
\affil[1]{Institute of Theoretical and Applied Informatics, Polish Academy of 
Sciences, Ba{\l}tycka 5, 44-100 Gliwice, Poland}
\affil[2]{Institute of Physics, Jagiellonian University, ulica prof.
Stanis\l{}awa \L{}ojasiewicza 11, Krak{\'o}w, Poland}

\date{28/12/2015}

\begin{document}

\maketitle

\begin{abstract}
In this work we aim at quantifying quantum channel output similarity. In order 
to achieve this, we introduce the notion of quantum channel superfidelity, 
which gives us an upper bound on the quantum channel fidelity. This quantity is 
expressed in a clear form using the Kraus representation of a quantum channel. 
As examples, we show potential applications of this quantity in the quantum 
control field.
\end{abstract}

\section{Introduction}\label{sec:introduction}
Recent applications of quantum mechanics are based on processing and
transferring information encoded in quantum states. The full description of
quantum information processing procedures is given in terms of quantum
channels, \textit{i.e.} completely positive, trace preserving maps on the set
of quantum states.

In many areas of quantum information processing one needs to quantify the
difference between ideal quantum procedure and the procedure which is performed
in the laboratory. This is especially true in the situation when one deals with
imperfections during the realization of experiments. These imperfections can be
countered, in a quantum control setup, using various techniques, such us
dynamical decoupling
\cite{viola1999dynamical,viola1999universal,viola2003robust,dahleh1990optimal},
sliding mode control~\cite{dong2009sliding} and risk sensitive quantum
control~\cite{james2004risk,d2006quantum}. A different approach is to model the
particular setup and optimize control pulses for a specific task in a specific
setup~\cite{pawela2013various,pawela2014quantum,gawron2014decoherence,pawela2014quantum2}.
In particular the problem of quantifying the distance between quantum channels 
was studied in the context of channel distinguishability.

One possible approach to quantifying the distance between two quantum channels
is to consider the fidelity between Choi-Jamio{\l}kowski states corresponding
to quantum channels~\cite{raginsky2001fidelity}. Another approach could involve
the diamond norm~\cite{piani2009all} of quantum channels. We propose an
approach which focuses on the outputs of quantum channels.

The main aim of this paper is to provide a succinct expression for the channel
output similarity. As a measure of similarity we will consider the
superfidelity function and define channel superfidelity. Then we will show
examples of application of our results to various pairs of quantum channels. In
the final part of the paper, we will study the impact of Hamiltonian errors on
the channel superfidelity. First we will consider a single qubit at a finite
temperature, and next we will move to an extended quantum control example.

\section{Preliminaries}\label{sec:preliminaries}

Henceforth we will denote the set of linear operators, transforming vectors
from a finite-dimensional Hilbert space $\XX$ to another finite dimensional
Hilbert space $\YY$ by $\LL(\XX, \YY)$. We put $\LL(\XX) = \LL(\XX, \XX)$. By
$\UU(\XX)$ we will denote the set of unitary operators on $\XX$. Given an
operator $A \in \LL(\XX, \YY)$ we denote by $\|A\|_p$ its Schatten p-norm. By
$\bar{A}$ we will denote the element-wise complex conjugation of $A$.

\subsection{Quantum states and channels}
First, we introduce two basic notions: density operators and superoperatros:
\begin{definition}
We call an operator $\rho \in \LL(\XX)$ a density operator iff $\rho \geq 0$
and $\tr \rho = 1$. We denote the set of all density operators on $\XX$ by
$\DD(\XX)$.
\end{definition}

From this follows that $\rho$ is in the form $\rho=\sum_j \lambda_j 
\proj{\lambda_j}$, where $\lambda_j$ and $\ket{\lambda_j}$ denote the 
$j^\mathrm{th}$ eigenvalue and eigenvector of $\rho$ respectively.

\begin{definition}
A superoperator is a linear mapping acting on linear operators $\LL(\XX)$ on a
finite dimensional Hilbert space $\XX$ and transforming them into operators on
another finite dimensional Hilbert space $\YY$. Thus
\begin{equation}
	\Phi: \LL(\XX) \rightarrow \LL(\YY).
\end{equation}
\end{definition}
Now we define the tensor product of superoperators
\begin{definition}
Given superoperators 
\begin{equation}
\Phi_1: \LL(\XX_1) \rightarrow \LL(\YY_1), \Phi_2: \LL(\XX_2) \rightarrow
\LL(\YY_2),
\end{equation}
we define the product superoperator
\begin{equation}
	\Phi_1 \otimes \Phi_2: \LL(\XX_1 \otimes\XX_2) 
	\rightarrow \LL(\YY_1\otimes \YY_2),
\end{equation}
to be the unique linear mapping that satisfies the equation
\begin{equation}
(\Phi_1 \otimes \Phi_2)(A_1 \otimes  A_2) = 
\Phi_1(A_1) \otimes \Phi_2(A_2),
\end{equation}
for all operators $A_1\in \LL(\XX_1), A_2 \in \LL(\XX_2)$.
\end{definition}

In the most general case, the evolution of a quantum system can be described
using the notion of a \emph{quantum
channel}~\cite{BZ2006,NC2000,puchala2011experimentally}.
\begin{definition}\label{def:channel}
A quantum channel is a superoperator $\Phi$ that satisfies the following restrictions:
\begin{enumerate}

\item $\Phi$ is trace-preserving, i.e. $\forall {A \in \LL(\XX)} \quad 
\tr(\Phi(A))=\tr(A)$,

\item \label{item:CP}$\Phi$ is completely positive, that is for every finite-dimensional
Hilbert space $\ZZ$ the product of $\Phi$ and identity mapping on $\LL(\ZZ)$ 
is a non-negativity preserving operation, i.e.
\begin{equation}
\forall {\ZZ} \ \forall {A \in \LL(\XX \otimes \ZZ)} \quad {A \geq 0} 
\Rightarrow \Phi\otimes 
\Id_{\LL(\ZZ)}(A) \geq 0.
\end{equation}
\end{enumerate}
\end{definition}

Many different representations of quantum channels can be chosen, depending on
the application. In this paper we will use only the Kraus representation.

\begin{definition}
The Kraus representation of a completely positive superoperator
(Def.~\ref{def:channel}\eqref{item:CP}) is given by a set of operators  $K_i
\in \LL(\XX, \YY)$. The action of the superoperator  $\Phi$ is given by:
\begin{equation}
\Phi(\rho)=\sum_i K_i \rho K_i^\dagger,
\end{equation}
\end{definition}
This form ensures that the sueroperator is completely positive. For it to be 
also trace-preserving we need to impose the following constraint on the Kraus 
operators
\begin{equation}
	\sum_i K_i^\dagger K_i=\Id_{\XX},
\end{equation}
where $\Id_{\XX}$ denotes the identity operator acting on the Hilbert space 
$\XX$.

\subsection{Superfidelity}
In this section we introduce the \emph{superfidelity}, along with its properties
\begin{definition}
Superfidelity of two density operators $\rho, \sigma \in \DD(\XX)$ is given by
\begin{equation}
G(\rho, \sigma) = \tr(\rho\sigma) + \sqrt{1 - \tr\rho^2}\sqrt{1 - \tr\sigma^2}.
\label{eq:superfidelity}
\end{equation}
\end{definition}
The superfidelity is an upper bound for the fidelity 
function~\cite{miszczak2009sub,BZ2006}.

Properties of the superfidelity~\cite{miszczak2009sub} $(\rho_1, \rho_2, 
\rho_3, 
\rho_4 \in \DD(\XX))$:
\begin{enumerate}
\item Bounds: $0 \leq G(\rho_1, \rho_2) \leq 1$.
\item Symmetry: $G(\rho_1, \rho_2) = G(\rho_2, \rho_1)$.
\item Unitary invariance: $G(\rho_1, \rho_2) = G(U \rho_1 U^\dagger, U \rho_2 
U^\dagger)$, where $U \in \UU(\XX)$.
\item Joint concavity~\cite{bound-tr-di}:
\begin{equation}
G (p \rho_1 + (1-p)\rho_2, p \rho_3 + (1-p)\rho_4) \leq pG(\rho_1, \rho_3) + 
(1-p) G(\rho_2, \rho_4)\label{eq:concavity}
\end{equation}
for $p \in [0, 1]$.
\item Super-multiplicavity:
\begin{equation}
G(\rho_1 \otimes \rho_2, \rho_3 \otimes \rho_4) \geq G(\rho_1, \rho_3) 
G(\rho_2, \rho_4).
\end{equation}
\item Bound for trace distance~\cite{puchala2009bound}
\begin{equation}
\frac12 \| \rho_1 - \rho_2 \|_1 \geq 1 - G(\rho_1, \rho_2).
\end{equation}
\end{enumerate}

\subsection{Supporting definitions}
In this section we define additional operations used in our proof. We begin 
with the \emph{partial trace}
\begin{definition}
For all operators $A, B$ the partial trace is a linear mapping  
defined as:
\begin{equation}
\tr_{\YY} A \otimes B = A \tr B.
\end{equation}
The extension to operators not in the tensor product form follows from 
linearity.
\end{definition}
We will also need the notion of conjugate superoperator
\begin{definition}
Given a quantum channel $\Phi: \LL(\XX) \rightarrow \LL(\YY)$, for every
operator $A \in \LL(\XX), B \in \LL(\YY)$ we define the conjugate superoperator
$\Phi^\dagger: \LL(\YY) \rightarrow \LL(\XX)$ as the mapping satisfying
\begin{equation}
\Tr (\Phi(A)B) = \Tr (A \Phi^\dagger(B)). \label{eq:conjugate-channel}
\end{equation}
\end{definition}
Note, that the conjugate to completely positive superoperator is completely 
positive, but is not necessarily trace-preserving

Next, we will define a reshaping operation, which preservers the 
lexicographical order and its inverse.
\begin{definition}
We define the linear mapping
\begin{equation}
\mathrm{res}: \LL(\XX, \YY) \rightarrow \YY \otimes \XX,
\end{equation}
for dyadic operators as
\begin{equation}
\mathrm{res}(\ketbra{\psi}{\phi}) = \ket{\psi}\overline{\ket{\phi}},
\end{equation}
for $\ket{\psi} \in \YY$ and $\ket{\phi} \in \XX$ and uniquely extended by linearity.
\end{definition}
We introduce the inverse of the $\mathrm{res}(\cdot)$
\begin{definition}
We define the linear mapping
\begin{equation}
\mathrm{unres}: \YY \otimes \XX \rightarrow \LL(\XX, \YY)
\end{equation}
such that
\begin{equation}
\forall X \in \LL(\XX, \YY) \quad \mathrm{unres}(\mathrm{res}(X)) = X.
\end{equation}
\end{definition}

\begin{remark}
For every choice of Hilbert spaces $\XX_1$, $\XX_2$, $\YY_1$ and $\YY_2$ and 
every choice of operators $A \in \LL(\XX_1, \YY_1)$, $B \in \LL(\XX_2, \YY_2)$ 
and $X \in \LL(\XX_2, \XX_1)$ it holds that:
\begin{equation}
(A \otimes B) \mathrm{res}(X) = \mathrm{res}(AXB^\mathrm{T})
\end{equation}
\end{remark}
\begin{remark}\label{remark:purification}
For any choice of Hilbert spaces $\XX$ and $\YY$ and any choice of $\ket{\zeta} 
\in \XX \otimes \YY$ and $A \in \LL(\YY, \XX)$ such that $\ket{\zeta} = 
\mathrm{res}(A)$ it holds that
\begin{equation}
\Tr_\YY \ketbra{\zeta}{\zeta} = \Tr_\YY(\mathrm{res}(A)\mathrm{res}(A)^\dagger)
= AA^\dagger.
\end{equation}
\end{remark}
Next, we introduce the \emph{purification} of quantum states:
\begin{definition}\label{def:purification}
Given Hilbert spaces $\XX$ and $\YY$, we will call $\ket{\zeta} \in \XX \otimes 
\YY$ a purification of $\rho \in \DD(\XX)$ if
\begin{equation}
\tr_{\YY}\proj{\zeta} = \rho.
\end{equation}
\end{definition}
\begin{theorem}\label{th:unitary-equiv}
For every choice of Hilbert spaces $\XX$ and $\YY$ and let $\ket{\phi}, 
\ket{\psi} \in \XX \otimes \YY$ satisfy
\begin{equation}
\Tr_\YY(\proj{\phi}) = \Tr_\YY (\proj{\psi}).
\end{equation}
Then there exists a unitary operator $U \in \UU(\XX)$ such that $\ket{\psi} = 
(\Id_\XX \otimes U) \ket{\phi}$
\end{theorem}
From Definition~\ref{def:purification}, Theorem~\ref{th:unitary-equiv} and 
Remark~\ref{remark:purification} we get that given a state $\rho \in \DD(\XX)$ 
its purification $\ket{\zeta} \in \XX \otimes \XX$ is given by:
\begin{equation}
\ket{\zeta} = (\Id_\XX \otimes U) \mathrm{res}(\sqrt{\rho}).
\end{equation}
Verification of this equation is straightforward. First, we note that we may 
omit the term ($\Id_\XX \otimes U)$. Next we apply 
Remark~\ref{remark:purification} which allows us to show that for this choice 
of $\ket{\zeta}$ we get:
\begin{equation}
\Tr_\YY \proj{\zeta} = \sqrt{\rho} \left( \sqrt{\rho} \right)^\dagger = \rho.
\end{equation}

\subsection{Quantum channel fidelity}
First, we introduce the \emph{fidelity} and \emph{channel 
fidelity}~\cite{raginsky2001fidelity}
\begin{definition}
Given two density operators $\rho,\sigma$ we define the fidelity 
between $\rho$ and $\sigma$ as:
\begin{equation}
F(\rho, \sigma) = \| \sqrt{\rho} \sqrt{\sigma} \|_1^2 = \left(\tr 
\sqrt{\sqrt{\rho} \sigma \sqrt{\rho}} \right)^2
\end{equation}
\end{definition}
\begin{definition}
Quantum channel fidelity of a channel $\Phi: \LL(\XX) \rightarrow \LL(\XX)$ for 
some $\sigma$ is defined as:
\begin{equation}
F_\mathrm{ch}(\Phi; \sigma) = \inf F(\xi, (\Phi \otimes \Id_{\LL(\ZZ)})(\xi)),
\label{eq:channel-fidelity}
\end{equation}
where the infimum is over all Hilbert spaces $\ZZ$ and all $\xi \in \DD(\XX 
\otimes \ZZ)$ such that $\tr_{\ZZ}\xi = \sigma$
\end{definition}
It can be shown~\cite{watrous2011notes} that this infimum is independent of 
$\xi$ and is given by
\begin{equation}
F_\mathrm{ch}(\Phi; \sigma) = \sum_i |\Tr(\sigma K_i)|^2,
\end{equation}
where $K_i$ form the Kraus representation of $\Phi$.

\section{Our results}
In this section we present our main theorem and its proof. In the second 
subsection we present a quantum circuit that allows one to measure the quantum 
channel superfidelity without performing full state tomography.

\subsection{{Theorem and proof}}
\begin{definition}
Consider two quantum channels $\Phi, \Psi: \LL(\XX) \rightarrow \LL(\XX)$ and a
density operator $\sigma \in \DD(\XX)$. We define the quantum channel
superfidelity to be:
\begin{equation}
G_\mathrm{ch}(\Phi, \Psi;\sigma )= \inf G((\Phi \otimes \Id_{\LL(\ZZ)})(\xi),
(\Psi \otimes \Id_{\LL(\ZZ)})(\xi)),\label{eq:ch-superfidelity}
\end{equation}
where the infimum is over all Hilbert spaces $\ZZ$ and over all purifications $\xi = \proj{\zeta} \in 
\DD(\XX \otimes \ZZ)$ of $\sigma$. 
\end{definition}
The channel superfidelity $G_\mathrm{ch}(\Phi, \Psi; \sigma)$ places a lower
bound on the output superfidelity of two quantum channels in the case of the
same input states. Henceforth, where unambigous, we will write the channel 
superfidelity as $G_\mathrm{ch}$.

\begin{theorem}\label{th:main-theorem}
Given quantum channels $\Phi, \Psi: \LL(\XX) \rightarrow \LL(\XX)$ with Kraus
forms given by the sets $\{ K_i \}_i$ and $\{ L_j \}_j$ respectively the 
quantum channel superfidelity is given by:
\begin{equation}
\begin{split}
G_\mathrm{ch} = &\sum_{i,j} |\tr \sigma K_i^{\dagger} L_j|^2 + \sqrt{1 -
\sum_{i,j} |\tr \sigma K_i^{\dagger} K_j|^2} \\ & \times\sqrt{1 - \sum_{i,j}
|\tr \sigma L_i^{\dagger} L_j|^2}.\label{eq:our-result}
\end{split}
\end{equation}
\end{theorem}

\begin{theproof}
As we limit ourselves only to pure states $\xi$, in order to calculate the 
superfidelity, we need to compute the following quantities:
\begin{equation}
\begin{split}
\tr \left(\Phi \otimes \Id_{\LL(\ZZ)} \right) (\proj{\zeta})
\left(\Psi \otimes \Id_{\LL(\ZZ)} \right) (\proj{\zeta}), \\
\tr \left[ \left(\Phi \otimes \Id_{\LL(\ZZ)} \right) (\proj{\zeta}) \right]^2,\\
\tr \left[ \left(\Psi \otimes \Id_{\LL(\ZZ)} \right) (\proj{\zeta}) \right]^2.
\end{split}
\label{eq:quantities}
\end{equation}
As the general idea is shared between all of these quantities, we will show 
here the calculation for the first one. We get
\begin{equation}
\begin{split}
\tr \left(\Phi \otimes \Id_{\LL(\ZZ)} \right) (\proj{\zeta})
\left(\Psi \otimes \Id_{\LL(\ZZ)} \right) (\proj{\zeta}) = \\
\tr \proj{\zeta} \left( \Phi^\dagger \circ \Psi \otimes \Id_{\LL(\ZZ)} 
\right)(\proj{\zeta}) =\\
\bra{\zeta} \left( \Phi^\dagger \circ \Psi \otimes \Id_{\LL(\ZZ)} \right) 
(\proj{\zeta})
\ket{\zeta},
\end{split}
\end{equation}
where the first equality follows from the definition of the conjugate
superoperator. The Kraus form of the suoperoperator $\Phi^\dagger \circ \Psi$
is given by the set $\{ K_i^\dagger L_j \}_{i, j}$. Now, we may write 
$\ket{\zeta}$ as
\begin{equation}
\ket{\zeta} = \mathrm{res}(\sqrt{\sigma} U),
\end{equation}
for some $U \in \LL(\ZZ, \XX)$ such that $UU^\dagger$ is a projector on the 
image of $\sigma$.
We obtain:
\begin{equation}
\begin{split}
\bra{\zeta} \left( \Phi^\dagger \circ \Psi \otimes \Id_{\LL(\ZZ)} \right) 
(\proj{\zeta})
\ket{\zeta} = \sum_{i,j} |\bra{\zeta} (K_i^{\dagger} L_j) \otimes 
\Id_{\ZZ} \ket{\zeta}|^2 = \\
\sum_{i,j} |\mathrm{res}({\sqrt{\sigma}U})^\dagger \left((K_i^{\dagger} L_j) 
\otimes \Id_{\ZZ} \right) \mathrm{res}({\sqrt{\sigma}U})|^2 = \\
\sum_{i,j} |\Tr U^\dagger \sqrt{\sigma} K_i^{\dagger} L_j \sqrt{\sigma} 
U|^2 = \sum_{i,j} |\tr \sigma K_i^{\dagger} L_j|^2.
\end{split}
\end{equation}
This quantity is independent of the particular purification of $\sigma$. 
Following the same path for the other two quantities shown in
Eq~\eqref{eq:quantities}, we recover the expression for the channel 
superfidelity from Eq~\eqref{eq:our-result}.
\end{theproof}

Since the superfidelity is an upper bound for the fidelity function, we obtain,
the following inequality:
\begin{equation}
G_\mathrm{ch}(\Phi, \Psi; \sigma) \geq 
F(\Phi, \Psi; \sigma),
\end{equation}
where $F(\Phi, \Psi; \sigma) =
\inf_{\ZZ, \xi} F((\Phi \otimes \Id_{\LL(\ZZ)})(\xi), (\Psi \otimes 
\Id_{\LL(\ZZ)})(\xi))$ 
and  the infimum is over all Hilbert spaces $\ZZ$ and all $\xi \in \DD(\XX 
\otimes \ZZ)$ such that $\tr_{\ZZ}\xi = \sigma$.

The following simple corollaries are easily derived from 
Theorem~\ref{th:main-theorem}.
\begin{corollary}
Given a quantum channel $\Phi$, the superfidelity between its input and output 
reduces to the channel fidelity of $\Phi$:
\begin{equation}
G_\mathrm{ch}(\Phi, \Id; \sigma) = F_\mathrm{ch}(\Phi; \sigma)
\end{equation}
\end{corollary}
\begin{theproof}
Assume $\Phi$ has the Kraus form $\{K_i\}_i$. Substituting the identity for 
$L_j$ in Eq.~\eqref{eq:ch-superfidelity} we recover 
Eq.~\eqref{eq:channel-fidelity} which completes the proof.
\end{theproof}

\begin{corollary}
If $\Phi$ is a unitary channel i. e. $\Phi(\rho) = U\rho U^\dagger$ for any $U 
\in \UU(\XX)$ and $\Psi': \rho \mapsto U^\dagger \Psi(\rho) U$, where $\Psi$ is 
an arbitrary quantum channel then
\begin{equation}
G_\mathrm{ch}(\Phi, \Psi; \sigma) = F_\mathrm{ch}(\Psi'; \sigma).
\end{equation}
\end{corollary}
\begin{theproof}
If $\Phi$ is a unitary channel, then the second term in
Eq.~\eqref{eq:our-result} vanishes. Let us assume that $\Psi$ has a Kraus form
$\{L_j\}_j$. We get $G_\mathrm{ch} = \sum_j |\Tr \sigma U^\dagger L_j|^2$.

The Kraus form of the channel $\Psi'$ is given by the set $\{ U^\dagger L_j:
K_j \in \LL(\XX) \}$. Using this in Eq.~\eqref{eq:channel-fidelity} we get
$F_\mathrm{ch}(\Psi'; \sigma) = \sum_j|\tr \sigma U^\dagger L_j|^2$. This
completes the proof.
\end{theproof}

\begin{corollary}
If $\sigma \in \DD(\XX)$ is a pure state i.e. $\sigma = \proj{\psi}$ then
\begin{equation}
G_\mathrm{ch}(\Phi, \Psi; \sigma) = G(\Phi(\sigma), \Psi(\sigma)).
\label{eq:cor-3}
\end{equation}
\end{corollary}

\begin{theproof}
Let us only focus on the first terms in Eq.~\eqref{eq:superfidelity} and 
Eq.~\eqref{eq:our-result}. We will denote these terms $T$ and $T_\mathrm{ch}$ 
respectively. Let us assume that channels $\Phi$ and $\Psi$ have Kraus forms 
$\{K_i\}_i$ and $\{L_j \}_j$ respectively. We get:
\begin{equation}
\begin{split}
T & = \tr \sum_{ij} K_i \proj{\psi} K_i^\dagger L_j \proj{\psi} L_j^\dagger \\
& = \sum_{ij} \bra{\psi} K_i^\dagger L_j \proj{\psi} L_j^\dagger K_i \ket{\psi} 
= \sum_{ij} |\bra{\psi} K_i^\dagger L_j \ket{\psi}|^2 = T_\mathrm{ch}.
\end{split}
\end{equation}
Performing similar calculations for other terms, we recover Eq.~\eqref{eq:cor-3}
\end{theproof}

\subsection{Quantum circuit for measuring channel superfidelity}
Using the quantum circuit shown in Fig.~\ref{fig:measure-circ} we can measure
the quantum channel superfidelity in an experimental setup. This setup allows
us to estimate the $\tr \left(\Phi \otimes \Id_{\LL(\ZZ)} \right)
(\proj{\zeta}) \left(\Psi \otimes \Id_{\LL(\ZZ)} \right) (\proj{\zeta}) = 2 p_0
- 1$, where $p_0$ is the probability of finding the top qubit in the state
$\ket{0}$. Modifying the circuit appropriately, we can measure all the
quantities shown in Eq.~\eqref{eq:quantities}.
\begin{figure}[htp!]
\[
\Qcircuit @C=2.8em @R0.7em {
	\lstick{\ket{0}}      & \gate{H}    & \ctrl{6} & \ctrl{8} & \gate{H}   
	& \meter\\
	&             &                     &            & \\
	& \gate{\Id}  & \qswap   & \qw      &\qw         & \qw\\
	\lstick{\ket{\zeta}} &             &                     &            
	& \\
	& \gate{\Psi} & \qw      &  \qswap  &\qw         & \qw\\
	&             &                     &            & \\
	& \gate{\Id}  & \qswap   & \qw      &\qw         & \qw\\
	\lstick{\ket{\zeta}} &             &                     &            
	& \\
	& \gate{\Phi} & \qw      & \qswap   &\qw         & \qw
}
\]
\caption{Quantum circuit for measuring $\tr \left(\Phi \otimes \Id_{\LL(\ZZ)}
\right) (\proj{\zeta}) \left(\Psi \otimes \Id_{\LL(\ZZ)} \right)
(\proj{\zeta})= 2 p_0 - 1$, where $p_0$ is the probability of finding the top
qubit in state $\ket{0}$ \cite{ekert02direct}. This allows direct estimation
of the channel superfidelity.} \label{fig:measure-circ}
\end{figure}

Note that this approach is far simpler, compared to estimating the channel 
fidelity which would require us to perform full state tomography. Furthermore, 
analytical calculations involving fidelity get cumbersome quickly, as it 
requires calculating expressions of the form $\| \sqrt{\Phi(\sigma)} 
\sqrt{\Psi(\sigma)}\|_1$.

\section{Simple examples}
In this section we provide a number of examples of the application of 
Theorem~\ref{th:main-theorem}.

\subsection{Erasure channel}
\begin{definition}
Given a quantum state $\xi \in \DD(\XX)$ the erasure channel is given by:
\begin{equation}
\Phi(A) = \xi,
\end{equation}
for any $A$ in $\LL(\XX)$. The Kraus form of this channel is given by the set
$\{ K_{ij}: K_{ij} = \sqrt{\lambda_i} \ketbra{\lambda_i}{j} \}_{ij}$, 
$\lambda_i$ and $\ket{\lambda_i}$ denote the $i$\textsuperscript{th} eigenvalue 
and the corresponding eigenvector of $\xi$.
\end{definition}

Let us consider the superfidelity between the erasure channel $\Phi$ and a
unitary channel $\Psi: \sigma \mapsto U\sigma U^\dagger$ for some $U \in
\UU(\XX)$. We note that the second term in Eq.~\eqref{eq:our-result} vanishes.
What remains is:
\begin{equation}
\begin{split}
G_\mathrm{ch} & = \sum_{ij} \lambda_i |\bra{j} \sigma U^\dagger 
\ket{\lambda_i}|^2 = \sum_j \bra{j} \sigma U^\dagger \left(  \sum_i \lambda_i 
\proj{\lambda_i} \right) U \sigma \ket{j} \\
& = \tr \sigma^2 \Psi^\dagger(\xi) \leq \sum_i \lambda_i^\downarrow 
\mu_i^\downarrow,
\end{split}
\end{equation}
where $\mu_i^\downarrow$ and $\lambda_i^\downarrow$ denote the eigenvalues of 
$\sigma$ and $\xi$ respectively, sorted in a descending order. The last 
inequality follows from von Neumann's trace inequality~\cite{horntopics}.

\subsection{Sensitivity to channel error}\label{sec:channel-error}
Consider a quantum channel $\Phi$ with the Kraus form $\{K_i \}$ and a quantum 
channel $\Psi: \rho \mapsto U_\epsilon \Phi(\rho)
U_\epsilon^\dagger$, where $U_\epsilon = \exp(- i \epsilon H) \in \UU(\XX)$. We 
get:
\begin{equation}
G_\mathrm{ch}(\Phi, \Psi; \sigma) = 1 + \sum_{ij} |\tr \sigma K_i^\dagger 
U_\epsilon K_j|^2 - \sum_{ij} |\tr \sigma K_i^\dagger 
K_j|^2.\label{eq:sensitivity-error}
\end{equation}
Now, we concentrate on the change of the quantum channel superfidelity under 
the change of $\epsilon$. As we are interested only in small values of 
$\epsilon$, we expand Eq.~\eqref{eq:sensitivity-error} up to the linear term in 
the Taylor series. For small values of $\epsilon$ we get:
\begin{equation}
G_\mathrm{ch} \approx 1 - 2\epsilon \sum_{ij} \Im \Tr \sigma K_i^\dagger H K_j 
\overline{\tr \sigma K_i^\dagger K_j}.
\end{equation}
Note that this depends on the value of the observable $H$ of the operator 
$K_j\sigma K_i^\dagger$.

\section{Sensitivity to Hamiltonian parameters}
In this section we will show how the channel superfidelity is affected by 
errors in the system Hamiltonian parameters. First, we will show analytical 
results for a single qubit system at a finite temperature. Next, we show 
numerical results for a simple, three-qubit spin chain.

\subsection{Single qubit at a finite temperature}
A single qubit at a finite temperature is described by the master equation
\begin{equation}
\begin{split}
\dot{\rho}(t) = -\ii \frac{\Omega + \epsilon}{2}[\sigma_z,\rho(t)] &+ \gamma_+
\left( \sigma_- \rho(t) \sigma_+ - \frac12 \{ \sigma_+ \sigma_-, \rho(t) \}
\right) +
\\
&+ \gamma_- \left( \sigma_+ \rho(t) \sigma_- - \frac12 \{ \sigma_- \sigma_+, 
\rho(t) \} \right),\label{eq:single-qubit}
\end{split}
\end{equation}
where $\sigma_+ = \ketbra{1}{0}$, $\sigma_- = \sigma_+^\dagger$ and $\epsilon$ 
is the error in $\Omega$. Our goal is to calculate the quantum channel 
superfidelity between the case when there is no error in $\Omega$, i.e. 
$\epsilon=0$ and the case with error in $\Omega$. Henceforth, we will assume 
$\gamma_- = \gamma_+ = 1$ for clarity.

For a given time $T$, Eq.~\eqref{eq:single-qubit} may be rewritten as
\begin{equation}
\rho(T) =  \Phi_T^\epsilon(\rho(0)),\label{eq:single-qubit-channel}
\end{equation}
where $\Phi_T^\epsilon$ is a quantum channel in the quantum dynamical
semigroup. A natural representation $M_{\Phi_T^\epsilon}$ for the channel
$\Phi_T^\epsilon$ may be found as~\cite{havel2003robust}:
\begin{equation}
M_{\Phi_T^\epsilon} = \ee^{-F T},
\end{equation}
where
\begin{equation}
\begin{split}
F = &- \ii \frac{\Omega + \epsilon}{2}(\Id \otimes \sigma_z - \sigma_z \otimes 
\Id) - \sigma_- \otimes \sigma_- - \sigma_+ \otimes \sigma_+ +\\
&+ \frac12 ( \sigma_+ \sigma_- \otimes \Id + \sigma_- \sigma_+ \otimes \Id + 
\Id 
\otimes \sigma_+ \sigma_- + \Id \otimes \sigma_- \sigma_+ ).
\end{split}
\end{equation}
In this representation we may rewrite Eq.~\eqref{eq:single-qubit-channel}  as
\begin{equation}
\mathrm{res}(\rho(T)) = M_{\Phi_T^\epsilon} \mathrm{res}(\rho(0)).
\end{equation}

The Choi-Jamio{\l}kowski representation of the channel $\Phi_T$ is given by
$D_{\Phi_T^\epsilon} = (M_{\Phi_T}^\epsilon)^\mathrm{R}$. Here, $M^\mathrm{R}$ 
denotes
the \emph{reshuffle} operation on matrix $M$~\cite{BZ2006}. Now, it is simple
to find the Kraus form of the channel $\Phi_T^\epsilon$. The Kraus operators 
are related
to the eigenvalues $\lambda_i$ and eigenvectors $\ket{\lambda_i}$ of
$D_{\Phi_T^\epsilon}$ in the following manner:
\begin{equation}
K_i^{\Phi_T^\epsilon} = \sqrt{\lambda_i} 
\mathrm{unres}(\ket{\lambda_i}).\label{eq:kraus-single-qubit}
\end{equation}
Inserting these Kraus operators into Eq.~\eqref{eq:our-result} we get
\begin{equation}
G_\mathrm{ch}(\Phi_T^0, \Phi_T^\epsilon; \rho) = 1 - 2 \ee^{-2T}(1 - \cos 
\epsilon T)\rho_{00}(0) \rho_{11}(0),
\end{equation}
where $\rho_{ii}(0) = \bra{i}\rho(0)\ket{i}$. Note that, we get $G_\mathrm{ch} 
= 1$ in two cases. First, for large $T$ and second when $\epsilon T = 
\frac{\pi}{2}$. As we are mainly interested in small values of $\epsilon$, we 
expand $\cos\epsilon T$ up to the second term in the Taylor series. We get:
\begin{equation}
G_\mathrm{ch}(\Phi_T^0, \Phi_T^\epsilon; \rho) \approx 1 - \epsilon^2T^2 
\ee^{-2T} \rho_{00}(0)\rho_{11}(0).
\end{equation}

In this setup the channel superfidelity has a quadratic dependence on the error 
parameter $\epsilon$. This should be compared with the results in 
Sec.~\ref{sec:channel-error}.

\subsection{Quantum control example}
In this section, we consider a three-qubit spin chain with dephasing 
interactions with the environment. We will consider piecewise constant control 
pulses. The time evolution of the system is governed 
by the equation:
\begin{equation}
\dot{\rho}(t) = -\ii [H, \rho(t)] + \gamma(\sigma_z \rho(t) \sigma_z - \rho(t)),
\end{equation}
where $H = H_\mathrm{d} + H_\mathrm{c}$. Here $H_\mathrm{d}$ is the drift term 
of the Hamiltonian given by
\begin{equation}
H_\mathrm{d} = J \sum_{i=1}^2 \sum_{\alpha \in \{x, y, z\}} 
\sigma_\alpha^i \sigma_\alpha^{i+1},
\end{equation}
where $\sigma_\alpha^i$ denotes $\sigma_\alpha$ acting on site $i$. We set the 
control Hamiltonian $H_\mathrm{c}$ to:
\begin{equation}
H_\mathrm{c} = \sum_{i}^N h_x(t_i) \sigma_x^1 + h_y(t_i) \sigma_y^1,
\end{equation}
where $h_x(t_i)$ and $h_y(t_i)$ denote the control pulses in the time interval 
$t_i$. We set the target to be
\begin{equation}
U_\mathrm{T} = \Id \otimes 1 \otimes \sigma_x,
\end{equation}
i.e. a NOT gate on the third qubit. We fixed the number of time intervals 
$N=64$, the total evolution time $T=6.1$ and the maximum amplitude of a single 
control pulse $\forall k \in {x, y} \; \max(|h_k|)=10$.

\begin{figure}[!h]
\centering
\subfloat[]{\includegraphics{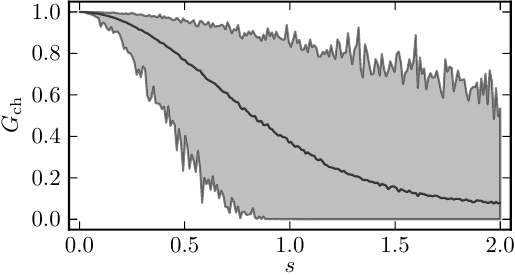}}\\
\subfloat[\label{fig:result-control-zoom}]{\includegraphics{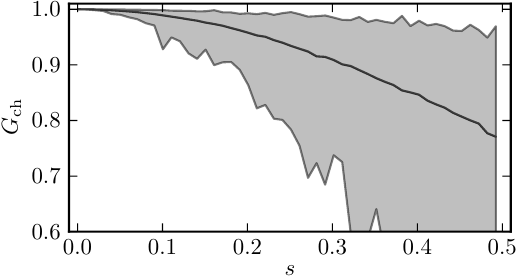}}
\caption{Quantum channel superfidelity as a function of noise in the system's 
control pulses. The shaded area represents the range of the achieved channel 
superfidelity, the black line is the average channel 
superfidelity.}\label{fig:result-control}
\end{figure}

First, we optimize control pulses for the system, such that we achieve a high 
fidelity of the gate $U_T$. Next, to each control pulse we add a noise term 
$h_\epsilon$ witch has a normal distribution, $h_\epsilon = N(0, s)$. 
Fig.~\ref{fig:result-control} shows the change of $G_\mathrm{ch}$ as a function 
of the standard deviation $s$. We have conducted 100 simulations for each 
value of $s$. As expected, the quantum channel superfidelity decreases 
slowly for low values of $s$. After a certain value the decrease becomes 
rapid. As values of $s$ increase, the minimum and maximum achieved 
fidelity diverge rapidly. This is represented by the shaded area in 
Fig.~\ref{fig:result-control}. We can approximate the average value of the 
channel fidelity as $\langle G_\mathrm{ch} \rangle \approx 1 - c s ^2$. Fitting 
this function to the curve shown in Fig.~\ref{fig:result-control-zoom} gives a 
relative error which is less then $0.5\%$.

\section{Conclusions}
We have studied the superfidelity of a quantum channel. This quantity allows us 
to provide an upper bound on the fidelity of the output of two quantum 
channels. We shown an example of application of this quantity to a unitary and 
an erasure channel. The obtained superfidelity can be easily limited from above 
by the product w eigenvalues of the input state $\sigma$ and the result of the 
erasure channel $\xi$.

Furthermore, as shown in our examples, the quantum channel superfidelity may
have potential applications in quantum control theory as an easy to compute
figure of merit of quantum operations. In a simple setup, where the desired 
quantum channel is changed by a unitary transformation $U_\epsilon = \exp(-\ii 
\epsilon H)$ we get a linear of the decrease of channel superfidelity on the 
noise parameter $\epsilon$. On the other hand, when we introduce the noise as a 
control error in a single qubit quantum control setup, we get a quadratic 
dependence on the noise parameter.

Finally, we shown numerical results for a more complicated system. We 
calculated  the quantum channel superfidelity for a three-qubit quantum control 
setup. First we found control pulses which achieve a high fidelity of the 
desired quantum operation, next we introduced Gaussian noise in the control 
pulses. Our results show, that the quantum channel superfidelity stayed high 
for a wide range of the noise strength.

\section*{Acknowledgements}
We would like to thank Piotr Gawron for inspiring discussions. {\L}P was
supported by the Polish National Science Centre under decision number
DEC-2012/05/N/ST7/01105. ZP supported by the Polish National Science Centre
under the post-doc programme, decision number DEC-2012/04/S/ST6/00400.

\bibliographystyle{ieeetr}
\bibliography{channel_superfidelity}

\begin{thebibliography}{10}

\bibitem{viola1999dynamical}
L.~Viola, E.~Knill, and S.~Lloyd, ``Dynamical decoupling of open quantum
  systems,'' {\em Physical Review Letters}, vol.~82, no.~12, p.~2417, 1999.

\bibitem{viola1999universal}
L.~Viola, S.~Lloyd, and E.~Knill, ``Universal control of decoupled quantum
  systems,'' {\em Physical Review Letters}, vol.~83, no.~23, p.~4888, 1999.

\bibitem{viola2003robust}
L.~Viola and E.~Knill, ``Robust dynamical decoupling of quantum systems with
  bounded controls,'' {\em Physical review letters}, vol.~90, no.~3, p.~037901,
  2003.

\bibitem{dahleh1990optimal}
M.~Dahleh, A.~Peirce, and H.~Rabitz, ``Optimal control of uncertain quantum
  systems,'' {\em Physical Review A}, vol.~42, no.~3, p.~1065, 1990.

\bibitem{dong2009sliding}
D.~Dong and I.~R. Petersen, ``Sliding mode control of quantum systems,'' {\em
  New Journal of Physics}, vol.~11, no.~10, p.~105033, 2009.

\bibitem{james2004risk}
M.~James, ``Risk-sensitive optimal control of quantum systems,'' {\em Physical
  Review A}, vol.~69, no.~3, p.~032108, 2004.

\bibitem{d2006quantum}
C.~D'Helon, A.~Doherty, M.~James, and S.~Wilson, ``Quantum risk-sensitive
  control,'' in {\em Decision and Control, 2006 45th IEEE Conference on},
  pp.~3132--3137, IEEE, 2006.

\bibitem{pawela2013various}
{\L}.~Pawela and P.~Sadowski, ``Various methods of optimizing control pulses
  for quantum systems with decoherence,'' {\em arXiv preprint arXiv:1310.2109},
  2013.

\bibitem{pawela2014quantum}
{\L}.~Pawela and Z.~Pucha{\l}a, ``Quantum control robust with respect to
  coupling with an external environment,'' {\em Quantum Information
  Processing}, vol.~14, no.~2, pp.~437--446, 2015.

\bibitem{gawron2014decoherence}
P.~Gawron, D.~Kurzyk, and {\L}.~Pawela, ``Decoherence effects in the quantum
  qubit flip game using markovian approximation,'' {\em Quantum Information
  Processing}, vol.~13, pp.~665--682, 2014.

\bibitem{pawela2014quantum2}
{\L}.~Pawela and Z.~Puchala, ``Quantum control with spectral constraints,''
  {\em Quantum Information Processing}, vol.~13, pp.~227--237, 2014.

\bibitem{BZ2006}
I.~Bengtsson and K.~{\.Z}yczkowski, {\em {Geometry of Quantum States: An
  Introduction to Quantum Entanglement}}.
\newblock Cambridge University Press, Cambridge, U.K., 2006.

\bibitem{NC2000}
M.~A. Nielsen and I.~L. Chuang, {\em {Quantum Computation and Quantum
  Information}}.
\newblock Cambridge University Press, 2000.

\bibitem{puchala2011experimentally}
Z.~Pucha{\l}a, J.~A. Miszczak, P.~Gawron, and B.~Gardas, ``Experimentally
  feasible measures of distance between quantum operations,'' {\em Quantum
  Information Processing}, vol.~10, no.~1, pp.~1--12, 2011.

\bibitem{miszczak2009sub}
J.~A. Miszczak, Z.~Pucha{\l}a, P.~Horodecki, A.~Uhlmann, and K.~Zyczkowski,
  ``Sub- and super-fidelity as bounds for quantum fidelity,'' {\em Quantum
  Information \& Computation}, vol.~9, no.~1, pp.~103--130, 2009.

\bibitem{bound-tr-di}
P.~E. Mendon{\c{c}}a, R.~d.~J. Napolitano, M.~A. Marchiolli, C.~J. Foster, and
  Y.-C. Liang, ``Alternative fidelity measure between quantum states,'' {\em
  Physical Review A}, vol.~78, no.~5, p.~052330, 2008.

\bibitem{puchala2009bound}
Z.~Pucha{\l}a and J.~A. Miszczak, ``Bound on trace distance based on
  superfidelity,'' {\em Physical Review A}, vol.~79, no.~2, p.~024302, 2009.

\bibitem{watrous2011notes}
J.~Watrous, {\em Theory of Quantum Information}.
\newblock 2011.

\bibitem{ekert02direct}
A.~K. Ekert, C.~M. Alves, D.~K.~L. Oi, M.~Horodecki, P.~Horodecki, and L.~C.
  Kwek, ``Direct estimations of linear and nonlinear functionals of a quantum
  state,'' {\em Phys. Rev. Lett.}, vol.~88, no.~21, p.~217901, 2002.

\bibitem{horntopics}
R.~Horn and C.~Johnson, ``Topics in matrix analysis, 1991,'' {\em Cambridge
  University Presss, Cambridge}.

\bibitem{havel2003robust}
T.~F. Havel, ``Robust procedures for converting among lindblad, kraus and
  matrix representations of quantum dynamical semigroups,'' {\em Journal of
  Mathematical Physics}, vol.~44, no.~2, pp.~534--557, 2003.

\end{thebibliography}
\end{document}